\documentclass[11pt,twoside]{article}

\usepackage{asp2006}
\usepackage{epsf}
\usepackage{psfig}
\usepackage{lscape}

\begin{document}
\title{Extreme Habitability: Formation of 
Habitable Planets in Systems with Close-in Giant Planets and/or 
Stellar Companions} 
\author{Nader Haghighipour}   
\affil{Institute for Astronomy and NASA Astrobiology Institute,
University of Hawaii-Manoa, Honolulu, Hawaii, USA}

\begin{abstract} 
With more than 260 extrasolar planetary systems discovered 
to-date, the search for habitable planets has found new grounds. 
Unlike our solar system, the stars of many of these planets are 
hosts to eccentric or close-in giant bodies. Several of these 
stars are also members of moderately close ($<$40 AU) binary or 
multi-star systems. The formation of terrestrial objects in 
these "extreme" environments is strongly affected by the dynamics
of their giant planets and/or their stellar companions. These 
objects have profound effects on the chemical structure of the 
disk of planetesimals and the radial mixing of these
bodies in the terrestrial regions of their host stars. For many 
years, it was believed that such effects would be so destructive 
that binary stars and also systems with close-in giant
planets would not be able to form and harbor habitable bodies. 
Recent simulations have, however, proven otherwise. I will review the 
results of the simulations of the formation and long-term stability of 
Earth-like objects in the habitable zones of such "extreme" planetary
systems, and discuss the possibility of the formation of terrestrial 
planets, with significant amounts of water, in systems with hot Jupiters, 
and also around the primaries of moderately eccentric close binary 
stars.

\end{abstract}

\section{Introduction}  
An analysis of the orbital properties of the currently known 
extrasolar planets indicates that many of these objects have 
dynamical characteristics that are profoundly different from
those of the planets in our solar system. While around the Sun, 
planets keep nearly circular orbits with smaller planets at 
closer distances and the larger ones farther way, from the 263 
exoplanets that were discovered at the time of the writing of 
this article, 68 were at distances smaller than 0.1 AU from 
their host stars, and 126 revolve in orbits with eccentricities 
larger than 0.2 (figure 1). A deeper look at the parent stars 
of these planets indicates that the differences between their 
planetary systems and the Solar System are not limited to the 
semimajor axes and eccentricities of their planetary companions. 
Several of these stars (approximately 20\% of currently known 
extrasolar planet-hosting stars) are members of binaries or
multi-star systems. 

Since the masses of the majority of extrasolar planets are within
the range of a few Neptune- to several Jupiter-masses, the 
eccentric and close-in orbits of these bodies, and the fact that
several of them exist around the components of binary stars,
raise questions on the formation of such {\it extreme} planetary 
systems, and also on the possibility of the existence of smaller 
planets, such as terrestrial-class objects, around their host 
stars, and particularly in their systems' habitable zones.
In this paper, I review the current status of research on the
formation of habitable planets in these extreme planetary environments,
and discuss the results of the simulations of Earth-like planet
formation in systems with close-in giant planets and binary
companions.

\begin{figure}
\plottwo {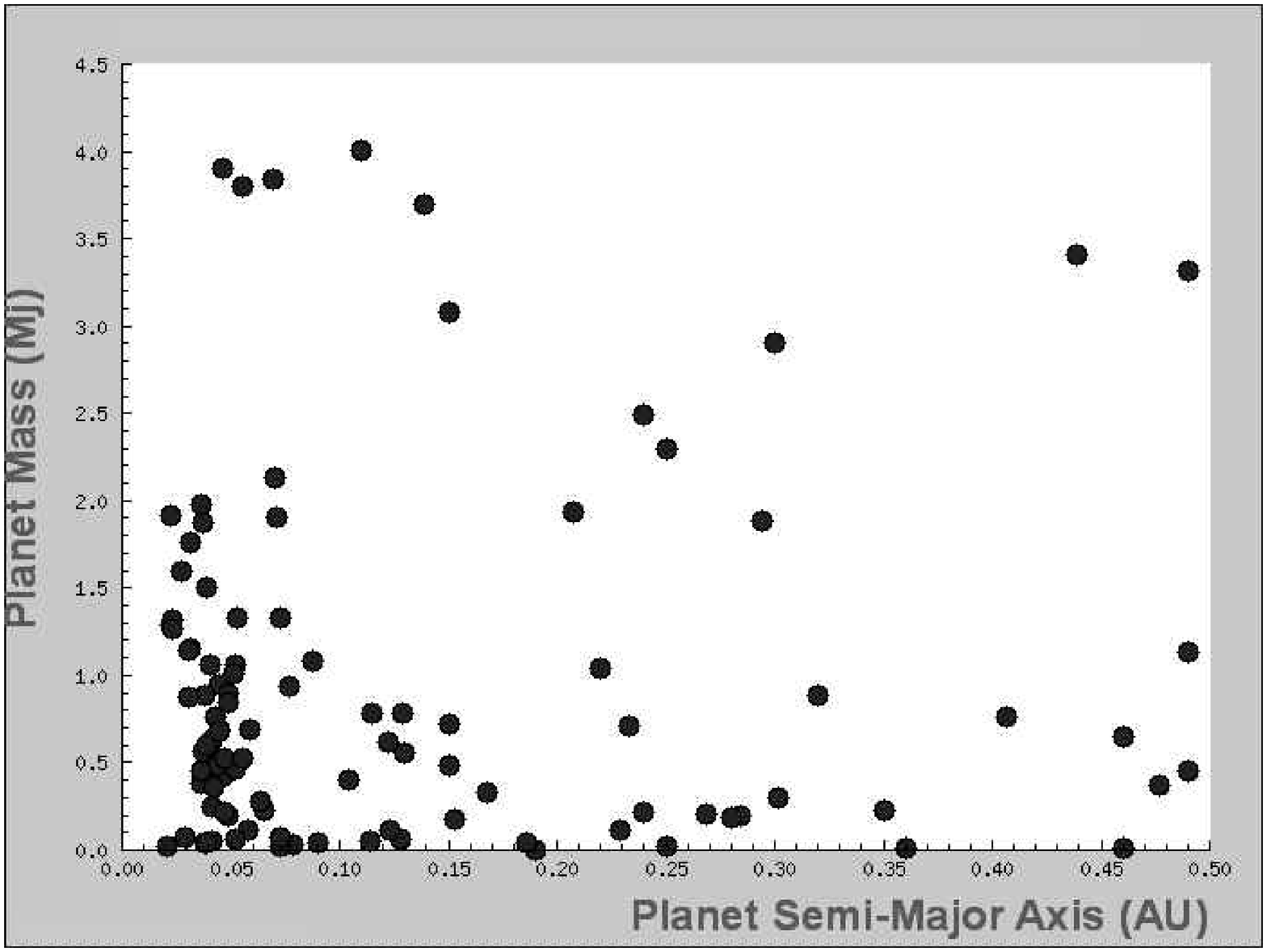}{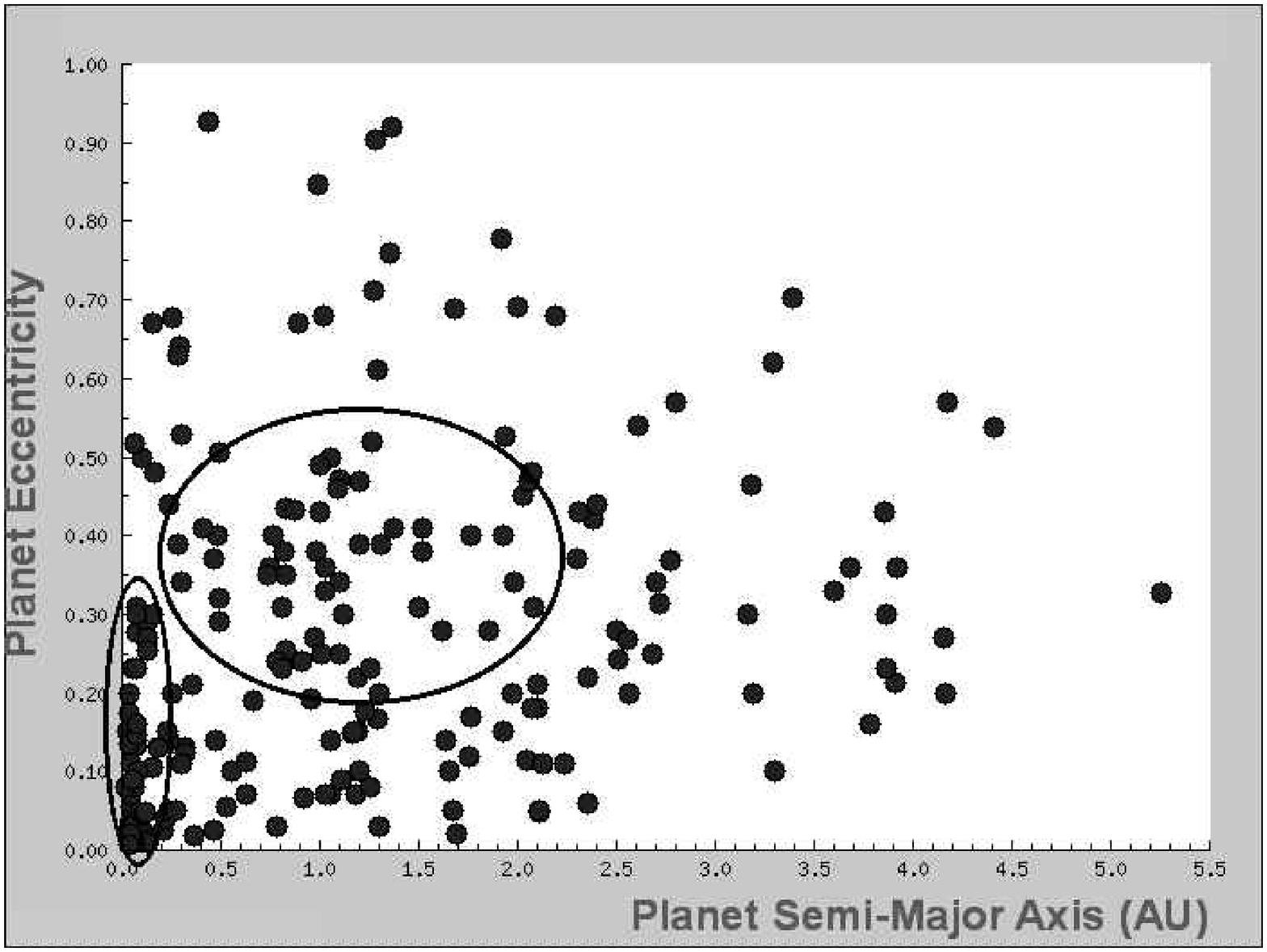}
\caption{Graphs of the mass and eccentricity of some of the
extrasolar planets. The graph on the left shows close-in giant
planets. The graph on the right shows planets with large eccentricities. 
The two encircled areas depict extreme planetary systems. Graphs from 
exoplanet.eu.}
\label{fig1}
\end{figure}

\section{Systems With Close-in Giant Planets}

It is widely accepted that the close-in giant planets have
formed at larger distances and have migrated to their current
positions. During their migrations, these objects perturb the orbits 
of smaller bodies on their paths and affect the interactions among
these objects. In general, the perturbative effect of a migrating
giant planet causes the orbital eccentricities and/or inclinations
of planetesimals and protoplanets to rise to higher values. Whether
these dynamically hot objects can cool down and re-interact
to form terrestrial bodies depends on the mass of the migrating
planet and its rate of migration. While, as shown in figure 2,
simulations by \citet{Armitage03} indicate that a giant planet 
migration will disrupt a disk of planetesimals and reduces its 
surface density to very low values (unless the migration occurs 
in a very short time), a systematic search of the parameter-space 
by \citet{Lufkin06} have resulted in identifying the ranges of the 
mass and rate of the migration of a giant planet for which 
dynamically excited planetesimals can return to low inclination 
and low eccentricity orbits (through gas-drag and dynamical friction) 
and increase the possibility of terrestrial planet formation (figure 2).
As shown by the four-panel graph in figure 2, for a low-mass 
(top left), or a fast migrating (bottom left) giant planet, the 
majority of planetesimals acquire eccentricities smaller than 0.4, 
implying that they may survive planetary migration and return to 
dynamically cool orbits in short times.

Recent numerical integrations by \citet{Fogg05,Fogg07},
\citet{Raymond06}, and \citet{Mandell07} have shown that
it is indeed possible for many planetesimals to survive giant
planet migration and form terrestrial bodies (figure 3).
As shown by \citet{Raymond06} and \citet{Mandell07}, 
\begin{figure}
\plottwo {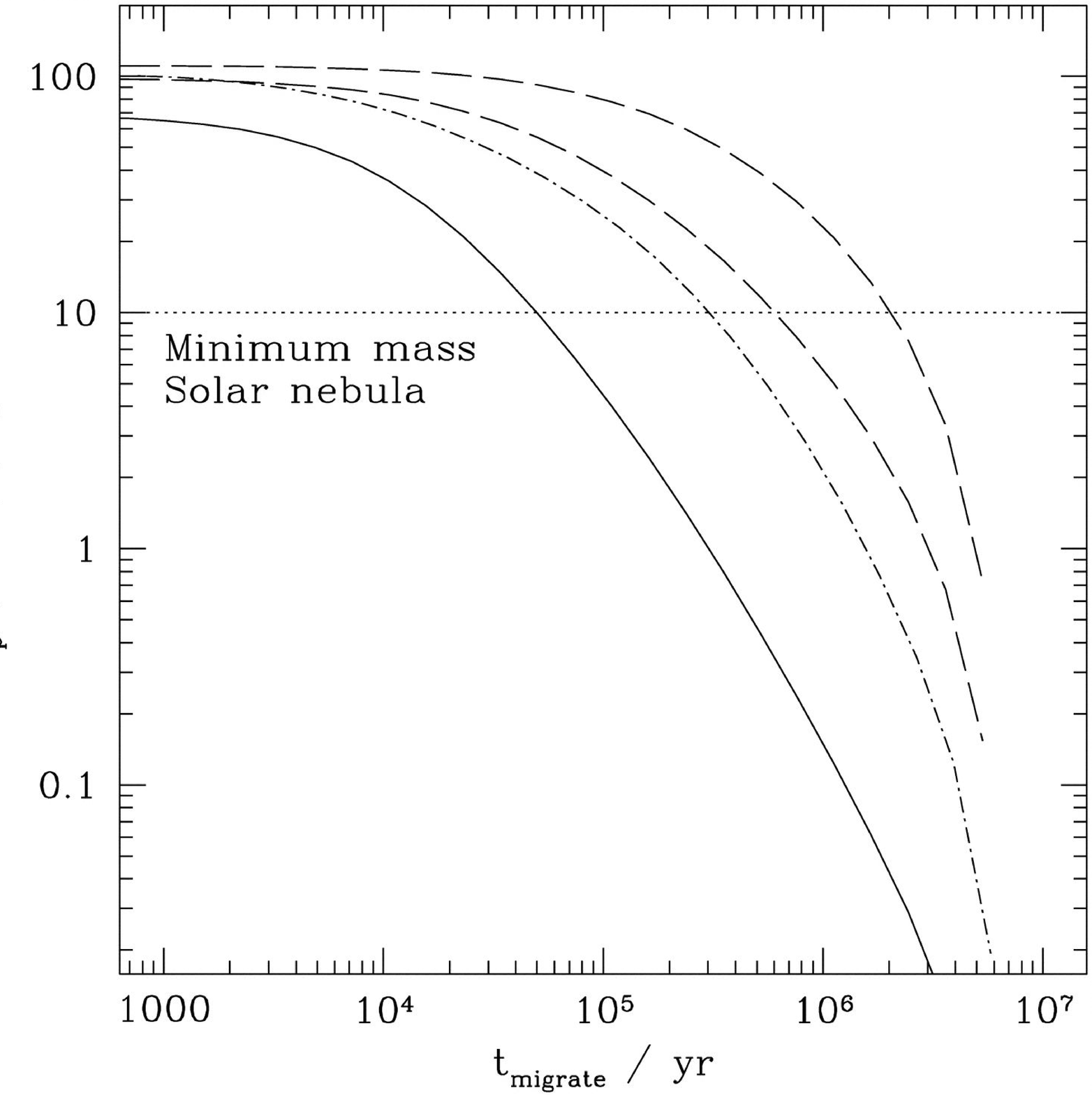}{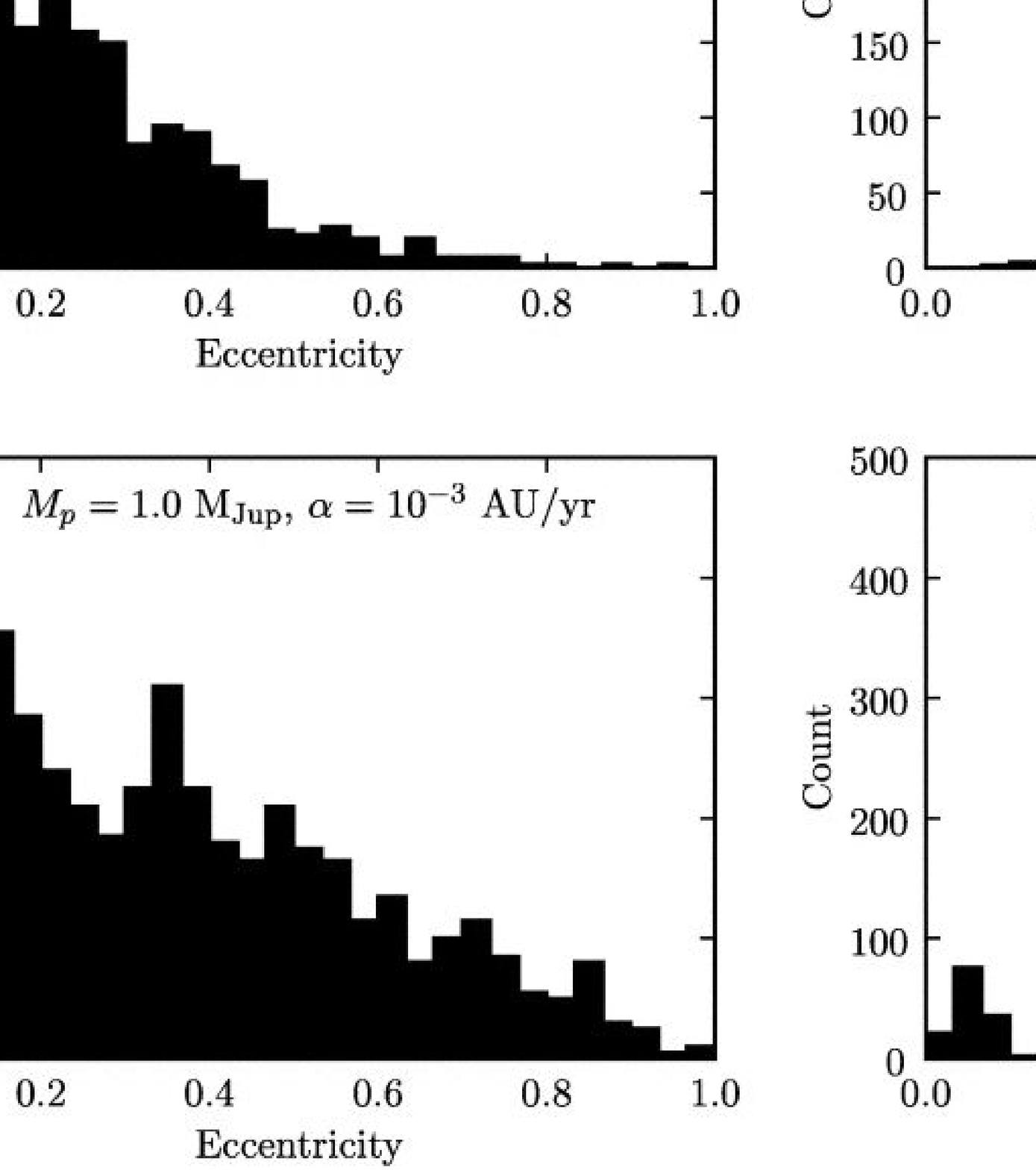}
\vskip -0.3in
\caption{Left: the variation of surface density of a disk of 
planetesimals in term of the time of the migration of a giant planet
\citep{Armitage03}. As shown here, unless the migration is fast,
giant planet migration destabilizes planetesimals resulting in
their removal from the disk. Right: eccentricity of planetesimals
during giant planet migration. As shown in the two left graphs,
for a low mass, or a fast migrating giant planet, the increase
in planetesimals' eccentricity may not be too high implying that
they may be able to return to low eccentricity orbits.}
\label{fig2}
\end{figure}
\noindent
under certain circumstances, not only do planetesimals survive 
giant planet migration, they can also form Earth-like objects, 
with substantial amounts of water, in the habitable zone of 
their central stars. By integrating the orbits of 1200 
planetesimals and 80 Moon- to Mars-sized objects distributed in 
a protoplanetary disk with a mass of 17 Earth-masses, these 
authors have shown that an Earth-like planet can form in the 
habitable zone of a star after 200 Myr, when a Jupiter-size object
migrates from 5 AU to 0.25 AU in $10^5$ years (figure 3). The
initial distribution of water in the disk of planetesimals is 
similar to those of our solar system's primitive asteroids, 
with a 0.5 water-to-mass ratio for protoplanetary objects beyond 
the location of the giant planet.

\vskip 20pt
\begin{figure}
\plottwo {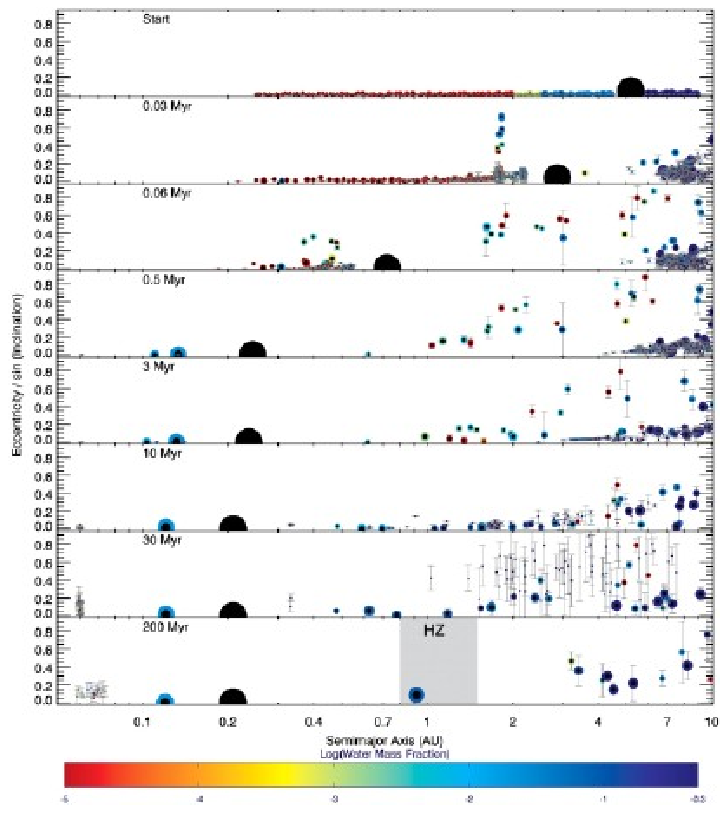}{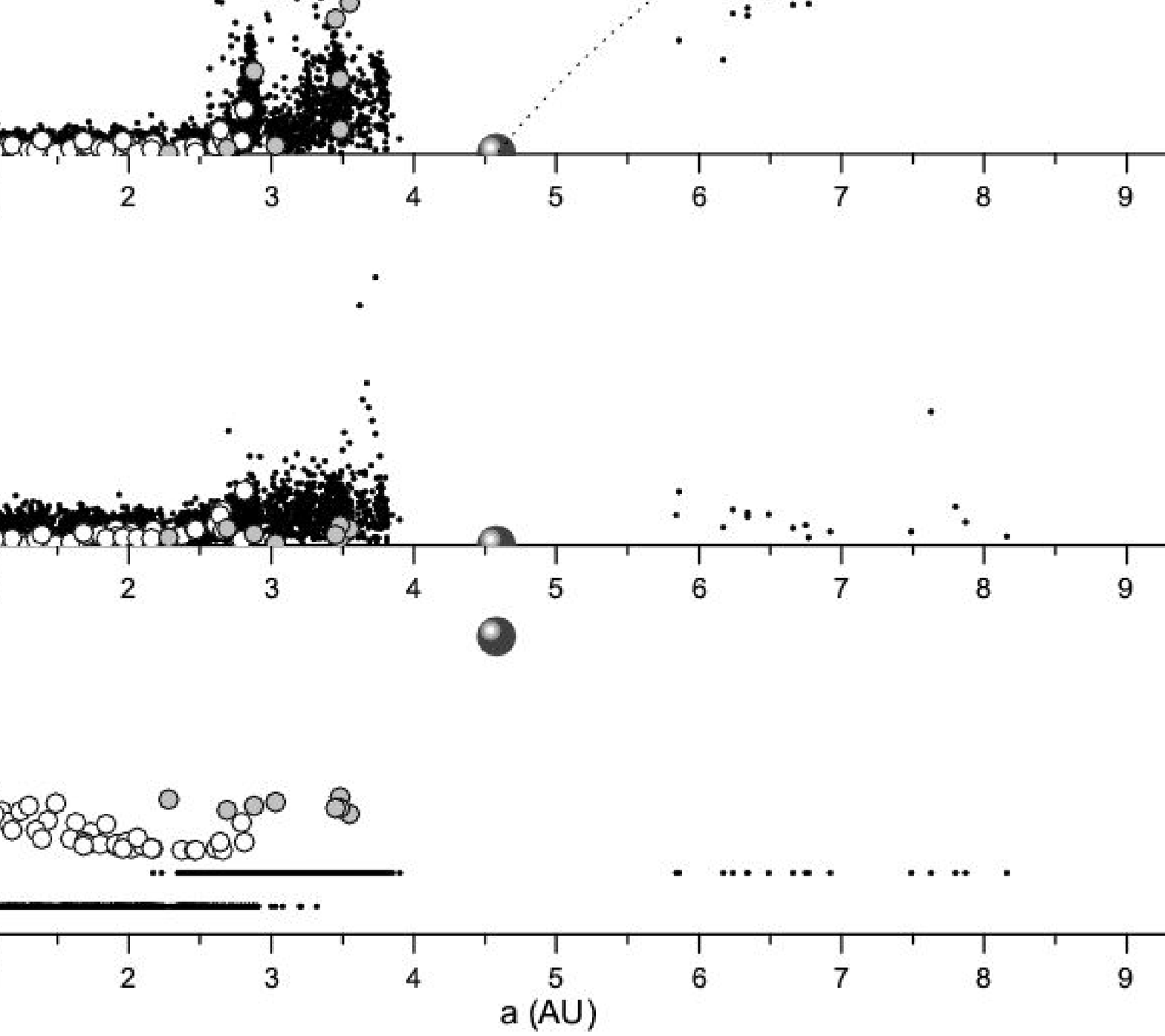}
\vskip -0.75in
\caption{Habitable planet formation during giant planet migration.
Graphs from \citet{Raymond06} left, and
\citet{Fogg07} right.}
\label{fig3}
\end{figure}
\noindent

\section{Habitable Planets in Binary Star Systems}

Given that a large fraction
of main and pre-main sequence stars are formed in binaries
or clusters \citep{Abt79, Duq91, Math00}, it is not
surprising that approximately 20\% of the currently known
extrasolar planet-hosting stars are members of binary systems. 
When the separation of these binaries are larger than 100 AU,
the perturbative effect of the gravitational force of one stellar
component on the formation and dynamics of planets around the
other star is negligible. However, when the separation of a 
binary is smaller than 100 AU, the perturbative effect of the 
farther companion becomes more pronounced, and strongly affects 
the long-term stability and the possibility of the formation of 
planets in binary systems.

Planet formation in binaries has been the subject of debate
for a long time. \citet{Artymowicz94} have shown that
a circumstellar disk around the 
\vfill
\hfill
\vskip -0.3in
\noindent
primary of a binary system may lose a large portion of its 
planet-forming material due to the interaction with an eccentric 
secondary star. \citet{Thebault04} have shown that such an 
interaction may also increase the orbital eccentricities of 
smaller objects inhibiting giant/terrestrial planet formation 
by destabilizing the orbits of their building blocks. However, 
the detection of circumstellar disks in systems such as L1551, 
in which the binary separation is $\sim$45 AU, and the stars of 
the system maintain circumstellar materials with masses of 
approximately 0.05 solar-masses (comparable to the minimum mass 
of our solar system's nebula) in disks with radii of 10 AU 
\citep{Rodriguez98}, and the recent detection of Jovian-type 
planets in three moderately close ($<$40AU) binaries of
$\gamma$ Cephei \citep{Hatzes03}, GL 86 \citep{Els01}, 
and HD 41004 \citep{Zucker04,Ragh06}, with separation smaller 
than 20 AU, have shown that it is possible for binary systems 
to maintain enough material in their circumstellar disks to 
trigger planet formation in the same fashion as around single 
stars. These {\it binary-planetary} systems, with their small 
separations, and with hosting Jupiter-like planets, present 
another case of extreme planetary systems. Whether such systems 
can harbor habitable planets depends on the degree of the 
interactions between their secondary stars, their giant planets, 
and their disks of embryos. As shown by \citet{Hagh06},
terrestrial-class objects in binary-planetary systems can maintain 
long-term stability in orbits close to their host stars and 
outside the influence zones of their systems' giant planets. 
That means,  in order for habitable planets to have stable 
orbits  in such systems, the habitable zones of their central 
stars have to be much closer to them than the orbits of their 
giant planets. In a recent article, we studied this topic by 
integrating the orbits of 120 Moon- to Mars-sized planetary embryos, 
with water contents similar to those of our solar system's 
primitive asteroids, in a system consisting of the Sun, Jupiter, 
and a farther stellar companion \citep{Hagh07}. Figure 4 shows 
the results for different values of the mass of the secondary 
star and its orbital parameters. As shown on the right graph, 
it is possible to form terrestrial-class objects, with substantial 
amounts of water, in the habitable zone of the primary star. 
The left graph of figure 4 shows a case in which a 1.17 Earth-masses
object is formed at 1.16 AU from the primary star, with an 
orbital eccentricity of 0.02, and water to mass ratio of 0.00164 
(Earth's water to mass ratio is approximately 0.001).   

The results of our simulations also indicate a relation between 
the perihelion of the binary $(q_b)$ and the semimajor axis of 
the outermost terrestrial planet $(a_{out})$. As shown in the 
left graph of figure 5, similar to  \citet{Quintana07},
simulations with no giant planets favor regions interior to 
$0.19{q_b}$ for the formation of terrestrial objects. That means, 
around a Sun-like star, where the inner edge of the habitable 
zone is at $\sim 0.9$ AU, a stellar companion with a perihelion 
distance smaller than 0.9/0.19 = 4.7 AU would not allow habitable 
planet formation. In simulations with giant planets, on the 
other hand, figure 5 shows that terrestrial planets form closer-in. 
The ratio ${a_{out}}/{q_b}$ in these systems is between 0.06 and 0.13.
A detailed analysis of our simulations also indicate that the 
systems, in which habitable planets were formed, have large 
perihelia. The right graph of figure 5 shows this for simulations
in a binary with equal-mass Sun-like stars. The circles in this 
figure represent systems with habitable planets.The numbers on 
the top of the circles show the mean eccentricity of the giant 
planet. For comparison, systems with unstable giant planets 
have also been marked. Since at the beginning of each
simulation, the orbit of the giant planet was considered to be 
circular, a non-zero eccentricity is indicative of the interaction 
of this body with the secondary star. As shown here, Earth-like 
objects are formed in systems where the interaction between the 
giant planet and the secondary star is weak and the  average 
eccentricity of the giant planet is small. That implies,
habitable planet formation is more favorable in binaries with 
moderate to large perihelia, and with giant planets on low 
eccentricity orbits.

\begin{figure}
\plottwo {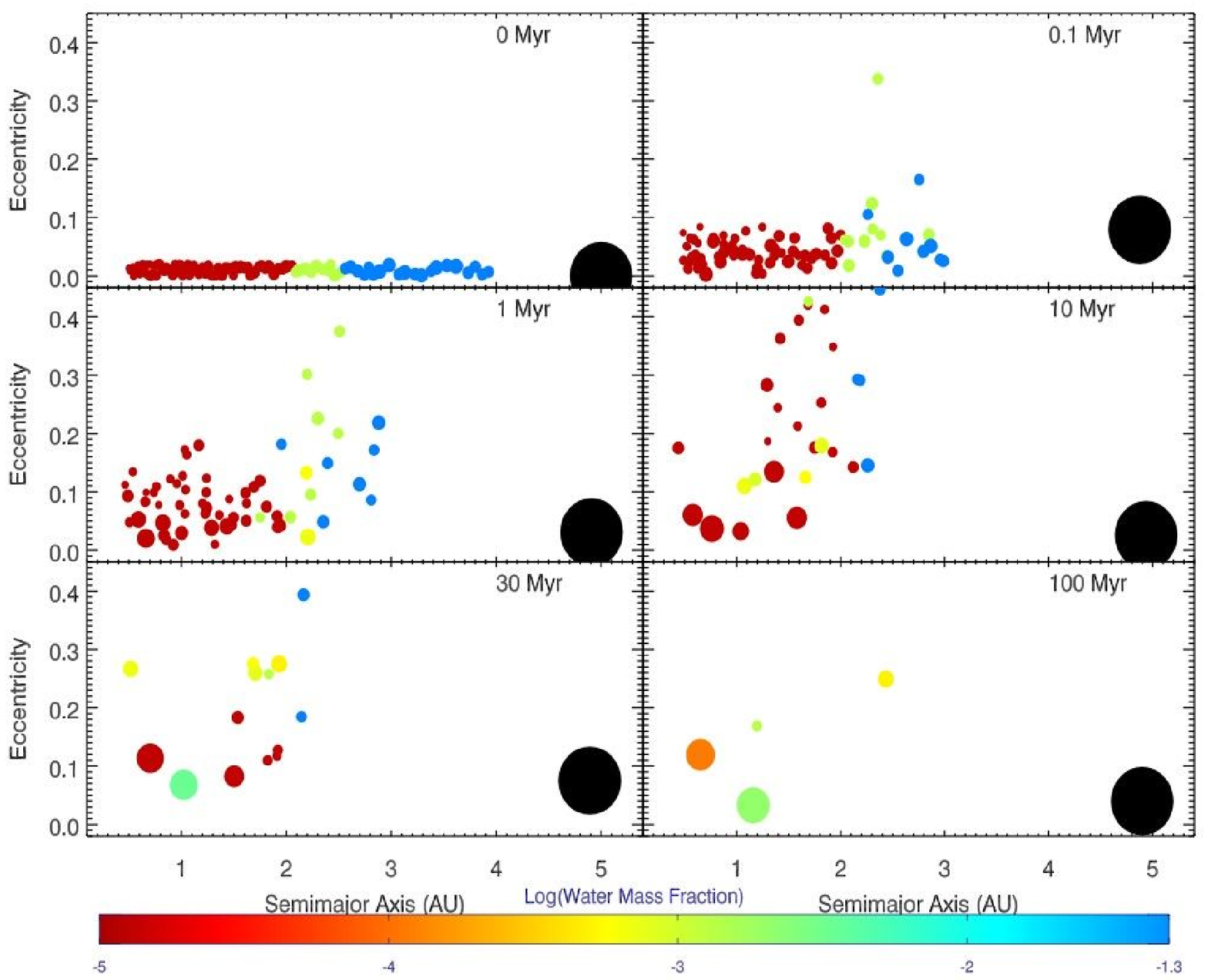}{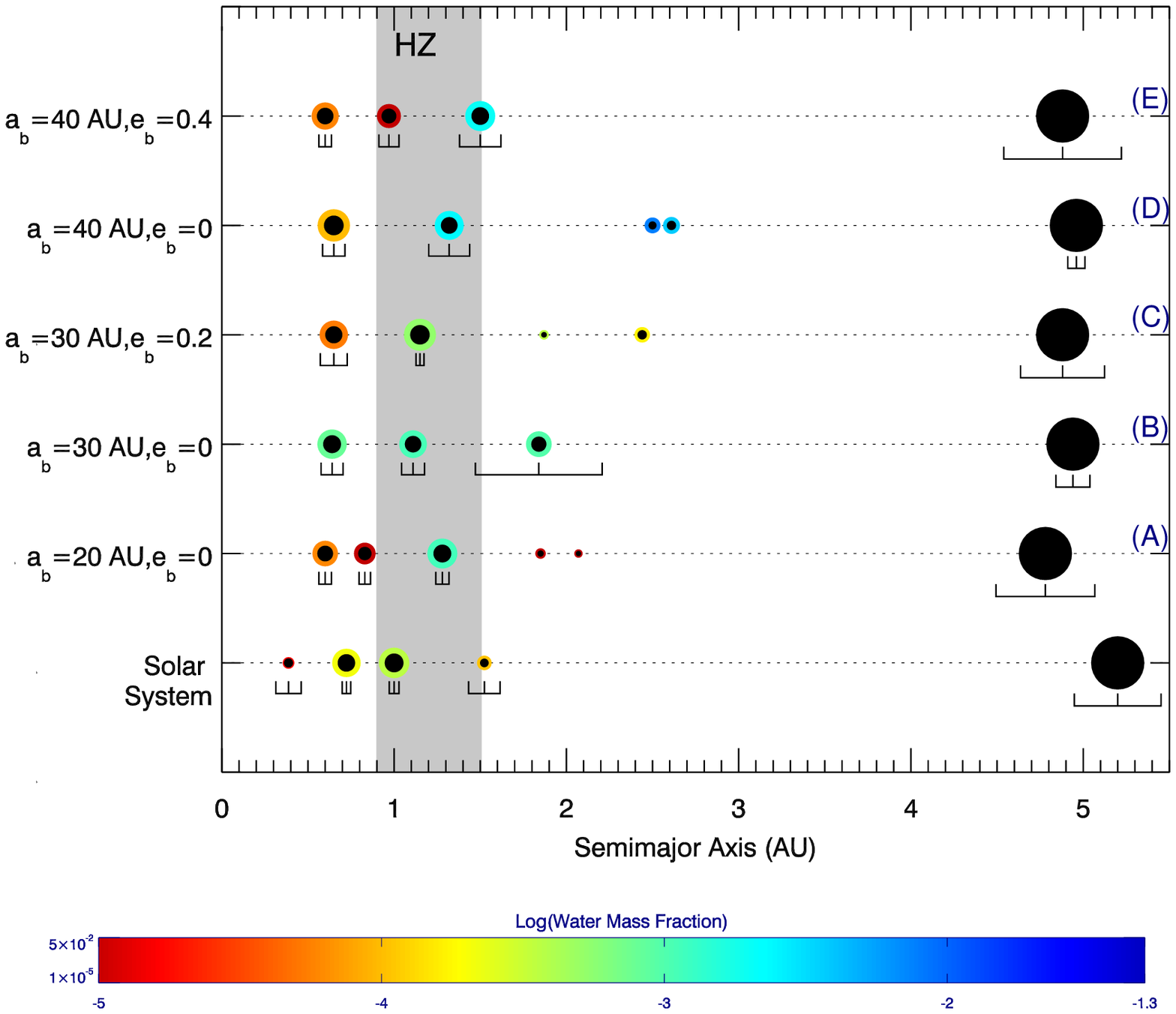}
\caption{Habitable planet formation in a binary star system with
Sun-like stars. Graph from \citet{Hagh07}.}
\label{fig4}
\end{figure}
\noindent

\begin{figure}
\plottwo {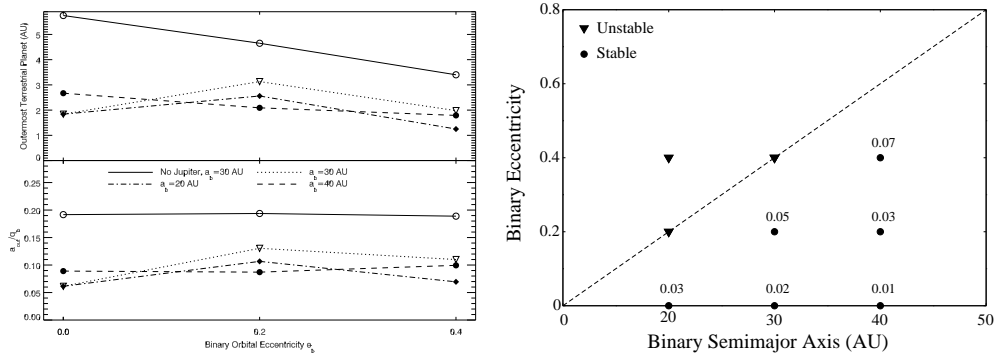}{f5b.eps}
\vskip 5pt
\caption{The graph on the left shows the relation between the 
perihelion of an equal-mass binary and the location of its outermost 
terrestrial planet. The graph on the right shows the region of 
the $({e_b},{a_b})$ space for a habitable binary-planetary system. 
Figures from \citet{Hagh07}.}
\label{fig5}
\end{figure}
\noindent

\vskip -2in
\acknowledgements 
Support for N.H. through the NASA Astrobiology Institute under 
Cooperative Agreement NNA04CC08A with the Institute for Astronomy 
at the University of Hawaii-Manoa is acknowledged.

\end{document}